%% file: main.tex
\documentclass{article}
\usepackage{graphicx} % Required for inserting images
\usepackage{lscape}

\usepackage[margin=1.5in]{geometry}    % For margin alignment
\usepackage[english]{babel}
\usepackage{tikz}
\usepackage{here}
\usetikzlibrary{positioning}
\usepackage{amsmath, amssymb}
\usepackage{booktabs}
\usepackage{hyperref}
\usepackage{multicol}
\usepackage{tikz}
\usetikzlibrary{quantikz2}
\usepackage{chemfig}
\usepackage{fancyvrb}
\usepackage{fvextra}
\title{Feasibility of first principles molecular dynamics in fault-tolerant quantum computer by quantum phase estimation}

\author{Ichio Kikuchi$^{1}$, Akihito Kikuchi$^{2}$\footnote{akihito\_kikuchi@gakushikai.jp (The corresponding author; a visiting researcher in IFQT)}  \\
        \small $^{1}$Internationales Forschungszentrum f\"ur Quantentechnik\\
        \small $^{2}$International Research Center for Quantum Technology, Tokyo \\
}

\date{\today}    % Today's date

\begin{document}\maketitle
\begin{abstract}
 This article shows a proof of concept regarding the feasibility of ab initio molecular simulation, wherein the wavefunctions and the positions of nuclei are simultaneously determined by the quantum algorithm, as is realized by the so-called Car-Parrinello method by classical computing. The approach used in this article is of a hybrid style, which shall be realized by future fault-tolerant quantum computer.  First, the basic equations are approximated by polynomials. Second, those polynomials are transformed to a specific form, wherein all variables (representing the wavefunctions and the atomic coordinates) are given by the transformations acting on a linear space of monomials with finite dimension, and the unknown variables could be determined as the eigenvalues of those transformation matrices. Third, the eigenvalues are determined by quantum phase estimation. Following these three steps, namely, symbolic, numeric, and quantum steps, we can determine the optimized electronic and atomic structures of molecules. 
\end{abstract}

\section{Introduction}
 
The authors of the present article have developed a method of quantum simulation of materials, wherein numeric, symbolic, and quantum algorithms are employed \cite{KIKUCHI2013, kikuchi2022abP, kikuchi2023molecular, kikuchi2023symbolic}. This approach uses the following steps.

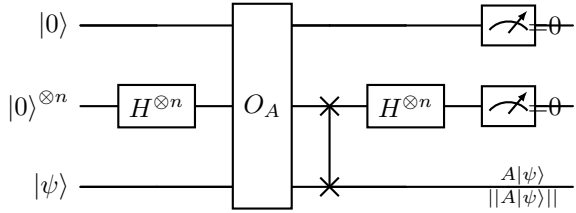
\begin{figure}[H]
\label{BLOCKENCODING}
\begin{center}
\begin{quantikz}
\lstick{$\ket{0}$}& \qw & \gate[3]{O_A} & \qw &\qw &\meter{}&\lstick{=0}\\
\lstick{$\ket{0}^{\otimes n}$}& \gate{H^{\otimes n}} & &\swap{1} & 
\gate{H^{\otimes n}} &\meter{}&\lstick{=0}\\
\lstick{$\ket{\psi}$}& \qw & \qw & \targX{} & \qw & &\lstick{$\frac{A\ket{\psi}}{||A\ket{\psi}||}$}
\end{quantikz}
\end{center}
\caption{
An illustration of the quantum circuit that executes block-encoding. The circuit is composed of Hadmard gates, the oracle $O_A$, and a swap gate. The query operation $O_A$ for a $2^n\times 2^n$ matrix $A=\{a_{ij}\}$ is given by
$ O_A |0\rangle|i\rangle |j\rangle =\left( a_{ij} \ket{0}+\sqrt{1-|a_{ij}|^2}\ket{1}\right) |i\rangle |j\rangle$. 
After the measurement, the quantum circuit generates 
$\frac{A\ket{\psi}}{||A\ket{\psi}||}$. 
 Note that the block-encoding is applied only for the matrix that satisfies $|a_{ij}|\le 1$ for all entries. If the matrix $A$ does not satisfy this condition, it should be multiplied by a suitable factor beforehand.
}
\end{figure}

\begin{itemize}
\item 
The molecular integrals are given by analytic functions generated from analytic atomic bases, such as GTO or STO, through the standard method of computational chemistry \cite{szabo2012modern,boys1950electronic}.
 
\item 
The energy is given by an analytic function. It is a multivariate polynomial, where the variables represent the coefficients of LCAO. In the same way, the ortho-normalization condition for the wavefunctions is prepared with Lagrange multipliers that represent orbital energies. If necessary, through the series expansion for other variables in the function, extra unknowns are included in the polynomial objective function for the total energy.
 
\item 
A system of polynomial equations is derived according to the minimum condition of the energy functional. 
\begin{align}
f_1=f_2=...=f_t=0
\end{align}
Those polynomials are represented by the coefficients of LCAO, the orbital energies, and other variables, say, the atomic coordinates. Henceforward, assume that those variables are represented by $x_1, x_2, ... $, and $x_n$.

\item 
The polynomials in the system of equations compose an ideal $I$ in a suitable commutative ring $R=C[x_1, x_2, ...,x_n]$, where $C$ is the number field. Then the Gr\"obner basis $G$ for $I$ can be computed \cite{eisenbud2013commutative, cox2013ideals, ene2011gr, decker2006computing, buchberger1965algorithm}. $G$ and $I$ are equivalent in such a way that the zeros of these two systems depict the same geometric object in the Cartesian coordinate space. 

\item 
An ideal $I$ in the commutative algebra could be decomposed as the intersection of primary ideals \cite{eisenbud2013commutative}:
\begin{equation*}
I = \bigcap_{i} p_i
\end{equation*}
Each primary ideal in the decomposition stands for a smaller subset of the original system of polynomial equations. By finding out the common zero set of the polynomials in each subset, we get a part of the roots of the originally given problem. Note that the decomposition of the ideal is a counterpart of the prime factorization of an integer: $n =p_1^{a_1} p_2^{a_2} \cdots p_n^{a_n}$.  

\item  The quotient ring $R/I(=R/G)$ should be zero-dimensional. Namely, the system of polynomial equations, represented by the corresponding Gr\"obner basis, has discrete roots.
 
%\item 
%In the quotient ring $R/G$, the variables $x_1, x_2, ...,x_n$ act as transformation matrices over the monomial basis of  $R/G$, when the latter are multiplied by the former. Those transformation matrices share the eigenvectors on account of their commutable properties.
\item We use the following algorithm \cite{sottile2002enumerative}. 

\item Let $\bar{x}_1,\bar{x}_2,...,\bar{x}_2$ 
be the representatives of $x_1,x_2,...,x_n$ in $R[x_1,x_2,...,x_n]/G$. Additionally, let $b$ be a vector that is composed of the representatives of the monomial basis of the quotient ring.
 
 \item For any $i$, the multiplication $\bar{x}_i \cdot b$ is represented by
 \begin{align}
 \bar{x}_i \cdot b  = b \cdot M_{x_i}
 \end{align}
 with a transformation matrix $M_{x_i}$. The entries of the matrix are numbers, but not symbols.
 
 \item As $M_{x_i}\cdot M_{x_j}=M_{x_j}\cdot M_{x_i}$, those transformation matrices share common eigenvectors $\{v_j |j=1,...,N_M\}$, where $N_m$ is size of the monomial basis $b$.
 
 \item Let us consider the eigenvalue problems, defined as follows,
 \begin{align}
 \bar{\xi}_i^{(j)} \ v_j  = v_j \cdot M_{x_i}
 \end{align}
 for $i=1,...,m$ and $j=1,...,N_M$. Those equations are solved numerically, and the eigenvalues give the common zero set of the polynomials included in the ideal $I$.  Namely, the eigenvalues give the roots of the set of polynomial equations in such a way that
 \begin{align}
f_1(\bar{\xi}_1^{(j)},\bar{\xi}_2^{(j)},...,\bar{\xi}_n^{(j)})
=f_2(\bar{\xi}_1^{(j)},\bar{\xi}_2^{(j)},...,\bar{\xi}_n^{(j)})=\cdots=f_t(\bar{\xi}_1^{(j)},\bar{\xi}_2^{(j)},...,\bar{\xi}_n^{(j)})=0
 \end{align}
 for $j=1,...,N_M$. Note that if eigenvectors $\{v_j\}_j$ for one of the $\{M_i\}_i$ is obtained, the other components of the roots are computed by
 \begin{align}
 \bar{\xi}_i^{(j)}= \frac{(v_j \cdot M_{x_i},v_j)}{ (v_j ,v_j)}.
 \end{align}
Each solution $(\bar{\xi}_1^{(j)},\bar{\xi}_2^{(j)},...,\bar{\xi}_n^{(j)})$  corresponds to a primary ideal $p_i$ in the primary ideal decomposition of the ideal $I=(f_1,...,f_t)=\bigcap_i p_i$.

\item 
The quantum phase estimation (QPE) computes the eigenvalues of those transformation matrices. The common eigenvector $\ket{v_j}$ of those matrices is encoded in the quantum states and the eigenvalues of the matrices $\{\bar{\xi}_l^{(j)} \}_l$ are successively recorded in the ancillary component. Namely, the result of the computation shall be encoded in a quantum state as follows:
\begin{align}
\ket{v_j}\ket{0}\ket{0}\cdots\ket{0} 
\xrightarrow[]{QPE}
\ket{v_j}\ket{\bar{\xi}_1^{(j)}}\ket{\bar{\xi}_2^{(j)}}\cdots\ket{\bar{\xi}_n^{(j)}}
\end{align}

As the transformation matrices are not Hermitian, their time evolution should be executed by special quantum circuits, wherein a matrix $A$ is embedded in the leading principal block of a larger unitary matrix $U$ acting on the full Hilbert space:
\begin{equation}
    U=\begin{pmatrix}
        A & *\\
        * & *\\
    \end{pmatrix}
\end{equation}
This sort of unitary operation is realized in a quantum circuit by the trick of block encoding \cite{camps2022fable}, as is illustrated in Figure \ref{BLOCKENCODING}. The block-encoding of $A$ in $U$ enables us to put an arbitrary state vector in the input and execute the matrix-vector multiplication: $A|v\rangle$. Consequently, the time-evolution $\exp(-\sqrt{-1}A)$ for the non-Hermitian matrix $A$, which is required in the QPE, is implemented in a quantum circuit.  

%The important point is how to construct the circuits for block-encoding as concisely as possible.
%The troublesome point in QPE with the block encoding is that the non-Hermitian matrices would have complex eigenvalues, which are useless in the current problem setting. To evaluate complex eigenvalues, one should apply extended versions of QPE proposed by several authors \cite{wang2010measurement, daskin2014universal,camps2022fable}. Or one might use the checking criterion to detect useless complex eigenvalues, as was discussed in \cite{kikuchi2023symbolic}.
\end{itemize}

In the next section, we show how to apply this computational scheme in the determination of any degrees of freedom involved in the problem, since the potential ability of this approach is not limited to the computation of wavefunction.

\section{The numerical experiment}
 
\subsection{The model description}
 
\begin{figure}[H]
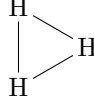

     \centering
\chemfig{H*3(-H-H-)}
     \caption{The molecular structure of H$_3^+$}
     \label{fig:molh3p}
 \end{figure}

We study a simple molecule H$_3^+$, as is shown in Figure \ref{fig:molh3p}. The shape of the molecule is an equilateral triangle with edge length $R$. Note that this molecule has attracted the interests of researchers in various fields since this molecule is the simplest triatomic molecule and abundant in the universe; this molecule is regarded as a benchmark problem of quantum chemistry  \cite{stevenson1937structure,hirschfelder1938energy,king1979theory,herbst2000astrochemistry,tashiro2002quantum,kokoouline2003theory,oka2005hot,foroutan2009chemical,pavanello2009calculate,jadhav2020theoretical,kannan2021rydberg}.  The electronic structure of this molecule as an algebraic variety is analyzed by the authors of the present works \cite{kikuchi2023molecular}. 

As a test problem, we pick up the simultaneous determination of the optimum electronic and atomic structures of this molecule. As the symbolic computation consumes computational resources, we simplify the problem by using the symmetric character of the molecule. Then the problem is reduced to the question of solving the model with three variables, which represent the essential elements in the molecular simulation, namely, a wavefunction, an orbital energy, and a bond length.

We adopt the following model.

\begin{itemize}
\item
To study the electronic structure of the molecule using molecular orbital theory, we use the STO-3G basis as the atomic orbital bases, from which we construct the energy function of the restricted Hartree-Fock model.  In the present case, we have only to consider only one molecular orbital. Namely, the degrees of the freedom of the system are given by the LCAO coefficients of three 1s orbitals $(x,y,z)$ the orbital energy $e$, and the bond length $R$. The necessary analytic formulas are given in the supplementary part of the present article. The symbolic computations to make up molecular integrals were carried out by a Python library {\it SymPy} \cite{10.7717/peerj-cs.103}.

\item
As we try to optimize the wave function and the atomic positions, the total energy is approximated by a truncated series representation (the Taylor expansion) for $R$. We choose the center of the expansion at $R_c=1.8\, a_0$ and take the terms up to the third order of $R$. 

\item
The coefficients in the total energy (with the constraint) $\{C_i\}_i$ are replaced by fractional numbers $\{D_i/10^n\}_i$  with a common denominator $10^n$. In this approximation, we set $n=8$. Then we multiply the total energy (with the constraint) by $10^n$ so that we use the objective function with integer coefficients. 

\item
We simulate the ground state of the molecule, by setting $x=y=z$. In other words, the atomic orbitals in the three atomic sites have equal LCAO coefficients.

\end{itemize}

With these conditions, the multivariate polynomial approximation of the objective function is given by:
 
\begin{Verbatim} [breaklines=true]
OBJ=-25940329*R**3*e*x**2 - 61451313*R**3*x**4 + 65640150*R**3*x**2 - 28577961*R**3 + 81961639*R**2*e*x**2 + 1099859207*R**2*x**4 - 811868595*R**2*x**2 + 205761316*R**2 + 342231572*R*e*x**2 - 5233649558*R*x**4 + 3595948148*R*x**2 - 555555556*R - 1960143305*e*x**2 + 200000000*e + 8467967598*x**4 - 6382868964*x**2 + 666666666
\end{Verbatim}

\subsection{The validity of the approximation}
 
\begin{figure}[H]
    \centering
    \includegraphics[width=0.8\textwidth, angle=0]{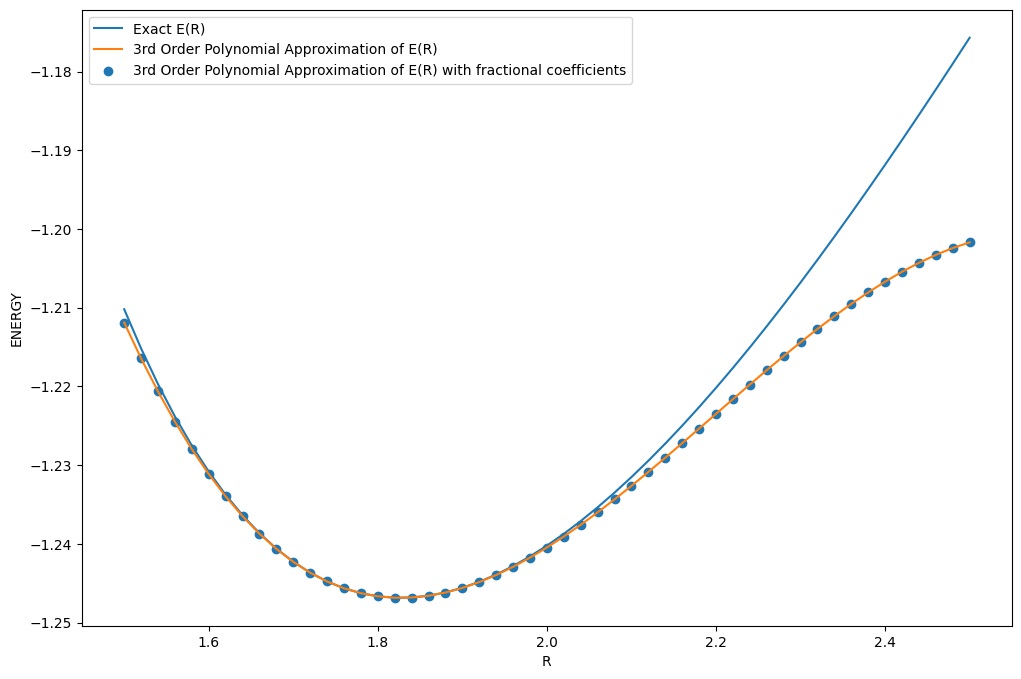}
    \caption{
    The total energy $E(R)$ and the bond length $R$ of H$_3^+$, computed by different levels of approximations: (1) the exact function (2) the polynomial approximation made by the series expansion for $R$ (3) the polynomial approximation with approximated coefficients given by fractional numbers.
    }
        \label{EnergyProfile}
\end{figure}

First, we check the validity of the polynomial approximation for the objective function. Figure \ref{EnergyProfile} shows the dependence of total energy on $R$ for three cases: (1) the exact energy function, (2) the polynomial approximation made by the series expansion for $R$, and (3) the polynomial approximation with the coefficients with fractional numbers, which is given in the present section.  Those three results coincide well at the neighborhood of $R_c=1.8$, around which the minimum of the energy functional is located.  

\subsection{Symbolic and numeric computation}
 
Second, we seek the optimum using numeric and symbolic algorithms. The optimum is given by the partial derivatives of the total energy with respect to $x$, $e$, and $R$.  In this step, we work in the ring $\mathbf{Q}[x,e, R]$, where the coefficient field is the rational number field.  The monomial ordering is first given by the degree reverse lexicographic type, and then it is switched to the lexicographic type. (In the last monomial ordering, the Gr\"obner basis could be decomposed into a set of triangular systems, and the roots of the given equations could be evaluated in each of those triangular systems. This type of numerical algorithm yields the same results as the eigenvalue approach that is demonstrated later.) The authors of the present article used the GAP system \cite{GAPSYSTEM} for symbolic computations related to Gr\"obner basis.

The entries of the polynomial ideal $I$ and the Gr\"obner basis $G$ are not given here, since they are lengthy and complicated.  The raw data are available via the internet. See the link given in Section \ref{DataAvailability}.  

The monomial basis $bs$ of the quotient field $\mathbf{Q}[x,e, R] / I$ is composed of the following elements. 

\begin{Verbatim} [breaklines=true]
bs={ xe4, ye4, x3eR, x2yeR, xy2eR, y3eR, xye2R, y2e2R, xe3R, ye3R, e4R, x3R2, x2yR2, xy2R2, y3R2, x2eR2, xyeR2, y2eR2, xe2R2, ye2R2, e3R2, xyR3, y2R3, xeR3, yeR3, e2R3, xR4, yR4, eR4, R5, x4, x3y, x2y2, xy3, y4, x3e, x2ye, xy2e, y3e, x2e2, xye2, y2e2, xe3, ye3, e4, x3R, x2yR, xy2R, y3R, x2eR, xyeR, y2eR, xe2R, ye2R, e3R, x2R2, xyR2, y2R2, xeR2, yeR2, e2R2, xR3, yR3, eR3, R4, x3, x2y, xy2, y3, x2e, xye, y2e, xe2, ye2, e3, x2R, xyR, y2R, xeR, yeR, e2R, xR2, yR2, eR2, R3, x2, xy, y2, xe, ye, e2, xR, yR, eR, R2, x, y, e, R, 1 }
\end{Verbatim}
 
The basis $bs$ generates a vector space of 22 dimensions.  The transformation matrices, which represent the multiplication of $x$, $y$, and $e$ over the monomial basis, are shown in the appendix. They are sparse, but not Hermitian.

%Let us solve the eigenvalue problem of $m_x$ and then compute the roots of the given system of the polynomial equations using the eigenvectors of $m_x$. The result is given in Table \ref{tab:my_labelxer}.

\begin{table}[H]
    \centering
\begin{tabular}{lccccc}
\toprule
{$\ket{i}$} &               x &                e &               R &           E$_{TOTAL}$ &     Type \\
\midrule
0  & -0.0000+0.2451j & 25.1507+0.0000j & -4.4726-0.0000j & 170.3683+0.0000j &  complex \\
1  & -0.0000-0.2451j & 25.1507-0.0000j & -4.4726+0.0000j & 170.3683-0.0000j &  complex \\
2  &  0.6014+0.0000j &  -1.8051+0.0000j & -3.8703+0.0000j &   8.1743+0.0000j &     real \\
3  & -0.6014+0.0000j &  -1.8051+0.0000j & -3.8703+0.0000j &   8.1743+0.0000j &     real \\
4  &  0.3703+0.0000j & -11.7442+0.0000j & -3.1022+0.0000j &  21.7662+0.0000j &     real \\
5  & -0.3703+0.0000j & -11.7442+0.0000j & -3.1022+0.0000j &  21.7662+0.0000j &     real \\
6  &  0.1137+0.1795j &  -0.6013-1.1800j &  0.1264-1.6890j &   0.9417+3.7454j &  complex \\
7  &  0.1137-0.1795j &  -0.6013+1.1800j &  0.1264+1.6890j &   0.9417-3.7454j &  complex \\
8  & -0.1137-0.1795j &  -0.6013-1.1800j &  0.1264-1.6890j &   0.9417+3.7454j &  complex \\
9  & -0.1137+0.1795j &  -0.6013+1.1800j &  0.1264+1.6890j &   0.9417-3.7454j &  complex \\
10 &  0.4050+0.0000j &  -1.1482+0.0000j &  1.8272+0.0000j &  -1.2469+0.0000j &     real \\
11 & -0.4050+0.0000j &  -1.1482+0.0000j &  1.8272+0.0000j &  -1.2469+0.0000j &     real \\
12 & -0.4580+0.0000j &  -0.8673+0.0000j &  2.6811+0.0000j &  -1.1895+0.0000j &     real \\
13 &  0.4580+0.0000j &  -0.8673+0.0000j &  2.6811+0.0000j &  -1.1895+0.0000j &     real \\
14 & -0.4790-0.0187j &  -0.9176-0.5313j &  2.8486-0.6587j &  -1.2421+0.0174j &  complex \\
15 & -0.4790+0.0187j &  -0.9176+0.5313j &  2.8486+0.6587j &  -1.2421-0.0174j &  complex \\
16 &  0.4790+0.0187j &  -0.9176-0.5313j &  2.8486-0.6587j &  -1.2421+0.0174j &  complex \\
17 &  0.4790-0.0187j &  -0.9176+0.5313j &  2.8486+0.6587j &  -1.2421-0.0174j &  complex \\
18 & -0.1775-0.3575j &  -0.5221+0.8834j &  2.9857-0.4501j &   1.6144+2.1018j &  complex \\
19 & -0.1775+0.3575j &  -0.5221-0.8834j &  2.9857+0.4501j &   1.6144-2.1018j &  complex \\
20 &  0.1775+0.3575j &  -0.5221+0.8834j &  2.9857-0.4501j &   1.6144+2.1018j &  complex \\
21 &  0.1775-0.3575j &  -0.5221-0.8834j &  2.9857+0.4501j &   1.6144-2.1018j &  complex \\
\bottomrule
\end{tabular}
    \caption{
The list of the solutions. Each row, indexed by $i=0,...,21$, shows the computed value of $x$, $e$, $R$, the total energy, and the type of the solutions (real or complex). We use the symbol $j$ to denote the imaginary unit.
    }
    \label{tab:my_labelxer}
\end{table}

In Table \ref{tab:my_labelxer}, the result of the computation is shown. The rows in the table show the sequential numbers of the normalized eigenvectors $|i)$ of the transformation matrix $m_x$ (for i=0,...,21), the expectation values $x=(i|m_x|i)$, $e=(i|m_e|i)$, $r=(i|m_r|i)$,  the corresponding total energy, and the type of the solutions (real or complex).

We should discard the solutions of complex values and adopt only the real solutions. Moreover, we should assess the real solutions. As the most hazardous approximation in the computation is the series expansion with respect to $R$, we should seek the solutions that fall in the proper range of this approximation.  We could guess that $|10)$ and $|11)$ would be appropriate solutions because $(10|m_r|10)$ and $(11|m_r|11)$ are located at the values closest to the center of the series expansion (given by $R_c=1.8$). Indeed they are the points where the energy functional takes the minimum value.

In this way, the atomic coordinates and the wavefunctions could simultaneously be determined in a run. The potential supremacy of the algorithm described here is that the present scheme does not require any iterative numerical computations to achieve the self-consistency and the optimum of the energy. 

As for the position of the center of the polynomial expansion, we can choose it in the range from 1.6 to 2.0 and we can reproduce the optimum almost equal to the one obtained in the above computation. 

\subsection{Quantum computation}
 
In this section, we discuss how to compute the roots of the given equation using quantum phase estimation.  We do not design the full circuit of the QPE, for the design of the QPE circuit is mere routine work. We focus on the construction of the quantum gate that carries out the time evolution since it is the key component of the algorithm.
 
As the matrices obtained in the present study are not Hermitian, we should use modified versions of QPE which could treat complex eigenvalues, as have been proposed by several authors \cite{wang2010measurement, daskin2014universal, shao2022computing}. These algorithms commonly use the trick of block-encoding, by which an arbitrary matrix $A$ is embedded in a larger unitary matrix $U$. In this section, we use the FABLE method \cite{camps2022fable} to implement the block-encoding of the transformation matrices $m_x$, $m_y$, and $m_z$ in quantum circuits.

Let us investigate the reliability of the block encoding in the present study.

Table \ref{tab:my_label} shows the accuracy of the block encoding. It is evaluated by the square norms of the difference between the transformation matrix $A$ and the top-left diagonal part $A_{BL}$ in the Hermitian matrix $U$ generated by the block-encoding: $\|A-A_{BL}\|^2$. The computed residues are almost zero.

\begin{table}[H]
    \centering
\begin{tabular}{lcc}
\toprule
{} &                $M$  &  $\|M-M_{BL}\|$ \\
\midrule
0 &                 $m_x$ &      1.938105e-24 \\
1 &                 $m_e$ &      7.236860e-22 \\
2 &                 $m_r$ &      2.597125e-22 \\
3 &  $\exp(-i\,m_x)$ &      1.800854e-24 \\
4 &  $\exp(-i\,m_y)$ &      3.391753e-21 \\
5 &  $\exp(-i\,m_r)$ &      3.794574e-21 \\
\bottomrule
\end{tabular}    \caption{
The accuracy of block-encoding for the transformation matrices $m_x$, $m_y$, and $m_z$. The difference between the matrix and the upper-left block of the unitary matrix of the corresponding block encoding are computed as the measures of the accuracy. 
}
    \label{tab:my_label}
\end{table}

Table \ref{tab:my_label2} shows the solutions of the optimization of the molecule H$_3^+$. In this case, instead of the transformation matrices used in the previous section, we used the emulation of the quantum circuit that conducts the block encoding. The result in this table agrees well with the result by the classical algorithm given in Table \ref{tab:my_label2})

\begin{table}[H]
    \centering
\begin{tabular}{lrrrrl}
\toprule
{$\ket{i}$} &               x &                e &               R &           ETOTAL &     Type \\
\midrule
0  & -0.0000-0.2451j &  25.1507-0.0000j & -4.4726-0.0000j & 170.3683-0.0000j &  complex \\
1  & -0.0000+0.2451j &  25.1507+0.0000j & -4.4726+0.0000j & 170.3683+0.0000j &  complex \\
2  &  0.6014-0.0000j &  -1.8051-0.0000j & -3.8703-0.0000j &   8.1743-0.0000j &     real \\
3  & -0.6014+0.0000j &  -1.8051-0.0000j & -3.8703-0.0000j &   8.1743-0.0000j &     real \\
4  & -0.3703+0.0000j & -11.7442-0.0000j & -3.1022-0.0000j &  21.7662+0.0000j &     real \\
5  &  0.3703-0.0000j & -11.7442-0.0000j & -3.1022-0.0000j &  21.7662+0.0000j &     real \\
6  & -0.1137-0.1795j &  -0.6013-1.1800j &  0.1264-1.6890j &   0.9417+3.7454j &  complex \\
7  & -0.1137+0.1795j &  -0.6013+1.1800j &  0.1264+1.6890j &   0.9417-3.7454j &  complex \\
8  &  0.1137+0.1795j &  -0.6013-1.1800j &  0.1264-1.6890j &   0.9417+3.7454j &  complex \\
9  &  0.1137-0.1795j &  -0.6013+1.1800j &  0.1264+1.6890j &   0.9417-3.7454j &  complex \\
10 & -0.4050+0.0000j &  -1.1482-0.0000j &  1.8272-0.0000j &  -1.2469-0.0000j &     real \\
11 &  0.4050-0.0000j &  -1.1482-0.0000j &  1.8272-0.0000j &  -1.2469+0.0000j &     real \\
12 & -0.4580+0.0000j &  -0.8673-0.0000j &  2.6811-0.0000j &  -1.1895+0.0000j &     real \\
13 &  0.4580-0.0000j &  -0.8673-0.0000j &  2.6811-0.0000j &  -1.1895+0.0000j &     real \\
14 & -0.4790+0.0187j &  -0.9176+0.5313j &  2.8486+0.6587j &  -1.2421-0.0174j &  complex \\
15 & -0.4790-0.0187j &  -0.9176-0.5313j &  2.8486-0.6587j &  -1.2421+0.0174j &  complex \\
16 &  0.4790-0.0187j &  -0.9176+0.5313j &  2.8486+0.6587j &  -1.2421-0.0174j &  complex \\
17 &  0.4790+0.0187j &  -0.9176-0.5313j &  2.8486-0.6587j &  -1.2421+0.0174j &  complex \\
18 & -0.1775-0.3575j &  -0.5221+0.8834j &  2.9857-0.4501j &   1.6144+2.1018j &  complex \\
19 & -0.1775+0.3575j &  -0.5221-0.8834j &  2.9857+0.4501j &   1.6144-2.1018j &  complex \\
20 &  0.1775+0.3575j &  -0.5221+0.8834j &  2.9857-0.4501j &   1.6144+2.1018j &  complex \\
21 &  0.1775-0.3575j &  -0.5221-0.8834j &  2.9857+0.4501j &   1.6144-2.1018j &  complex \\
\bottomrule
\end{tabular}
    \caption{
The list of the solutions computed from the block-encoded matrices. The solutions are presented in the same way as in Table \ref{tab:my_labelxer}. 
}
    \label{tab:my_label2}
\end{table}

Table \ref{tab:my_labelexp} shows the check of the accuracy of the block encoding of the time-evolution operators ($\exp(-1jA)$) for matrices $A$, by computing the expectation values of $\exp(-1jA)$. The block encoding yields quantitatively accurate results.

\begin{table}[H]
    \centering
\begin{tabular}{lrr}
\toprule
Operator & $\ket{10}$               & $\ket{11}$                \\
\midrule
$m_x$ (Block Encoding)                 &  0.4050-0.0000j & -0.4050+0.0000j \\
$m_e$ (Block Encoding)                 & -1.1482-0.0000j & -1.1482-0.0000j \\
$m_r$ (Block Encoding)                 &  1.8272-0.0000j &  1.8272-0.0000j \\
$\exp(-i\,m_x)$ (Block Encoding)         &  0.9191-0.3940j &  0.9191+0.3940j \\
$\exp(-i\,m_e)$ (Block Encoding)         &  0.4102+0.9120j &  0.4102+0.9120j \\
$\exp(-i\,m_r)$ (Block Encoding)         & -0.2536-0.9673j & -0.2536-0.9673j \\
$\exp(-i\,x)$ (Direct) &  0.9191-0.3940j &  0.9191+0.3940j \\
$\exp(-i\,e)$ (Direct) &  0.4102+0.9120j &  0.4102+0.9120j \\
$\exp(-i\,R)$ (Direct) & -0.2536-0.9673j & -0.2536-0.9673j \\
\bottomrule
\end{tabular}
    \caption{
The accuracy of the block encoding for time evolution operators. The columns in this table show the expectation values of the operators $m_x$, $m_e$, $m_r$, $\exp(-i\,m_x)$, $\exp(-i\,m_e)$, and $\exp(-i\,m_r)$ by the eigenvectors $\ket{10}$ and $\ket{11}$ in Table \ref{tab:my_labelxer}. The results are computed by two modes: (A) the block encoding and (B) the direct substitution of actual values of $x$, $e$, and $R$ to $\exp(\cdot)$. The results by block-encoding and the direct substitution are distinguished by the signs (Block Encoding) and (Direct). The results of these two modes of the computations agree well.
    }
    \label{tab:my_labelexp}
\end{table}

The numerical experiments given in Table \ref{tab:my_label}, Table \ref{tab:my_label2}, and Table \ref{tab:my_labelexp} affirm the accuracy of the block-encoding. We could determine the wavefunction and the atomic coordinates simultaneously by the quantum algorithm, provided that the given problem is transformed into a suitable form for quantum computations. To detect the optimum solution,  we could use the same check for the computed value of $R$, as was used in the computation by the classical algorithm in the previous section. 

%Figure: Quantum circuit
 
%Table: Gates included in the quantum circuit

%Table: The accuracy of the quantum computation by block-encoding
 
%These results show that the quantum circuits made by the FABLE method have achieved trustworthy efficiency. 

%The Table … shows the number of quantum gates in those quantum circuits. The result shows that the FABLE method is successful in generating concise quantum circuits. A primitive implementation for a $N \times N$ matrix A requires $N \times N$ rotation gates \cite{}.  For the case of $m_x$, $m_y$, and $m_z$, the number of the controlled rotation gates amounts to $22\times22$. In contrast, the data in Table *** shows that the quantum circuits made by FABLE methods use fewer rotation gates. 

To conclude this section, let us briefly review how the algorithm to evaluate complex eigenvalues works, using the method proposed in \cite{daskin2014universal}. It is an extension to the standard QPE \cite{PhysRevA.76.030306}. We assume the following conditions: the eigenvector $\ket{\psi_j}$  for the operator $U=\exp(2i \pi A)$ has the eigenvalue $\lambda_j=|\lambda_j|e^{i2\pi\phi}$, such that $|\lambda_j| \leq 1$; the binary representation of $\phi$ is given by $(0.x_1x_2\cdots x_m)$. We could fulfill the first requirement $|\lambda_j| \leq 1$ by multiplying a scale factor with the matrix $A$ under investigation.

\begin{itemize}
\item
We give the initial input of the iterative QPE as
\begin{align}
\ket{\Psi}_i=\ket{0}_p\ket{0\cdots 0}_a\ket{\psi_j}
\end{align}

\item
To determine a digit of $\phi$ ($x_k$), after several gate operations, we get the following state:
\begin{align}
\ket{\Psi}_{f} \sim \left( (1+e^{i2\pi(0.x_k)}|\lambda|^{2^k})\ket{0}_p +(1-e^{i2\pi(0.x_k)}|\lambda|^{2^k}))\ket{1}_p\right)\ket{0\cdots 0}_a\ket{\psi_j}
\end{align}

\item
The measurement of the amplitudes on $\ket{0}_p$ and $\ket{1}_p$ enables us to determine $x_k$ and $|\lambda|$, thanks to the following facts.

If $x_k=1$, then
\begin{align}
\ket{\Psi}_f \sim\left(
\left(1-|\bar{\lambda}|^{2^k})\right)\ket{0}_p+
\left(1+|\bar{\lambda}|^{2^k})\right)\ket{1}_p
\right)\ket{0\cdots 0}_a\ket{\psi_j}
\end{align}
In this case, since $|1-|\lambda|^{2k} |< |1+|\lambda|^{2^k}|$, 
the probability to observe the state $\ket{1}$ is larger. 

If $x_k=0$, then
\begin{align}
\ket{\Psi}_f\sim \left(
\left(1+|\bar{\lambda}|^{2^k})\right)\ket{0}+
\left(1-|\bar{\lambda}|^{2^k})\right)\ket{1}
\right)\ket{0\cdots 0}_a\ket{\psi_j}
\end{align}
In this case, the probability of observing the state $\ket{0}$ is larger.

In both cases, we can evaluate $|\lambda|^{2^k}$ from the measured amplitudes of $\ket{0}_p$ and $\ket{1}_p$. As is pointed out in \cite{daskin2014universal}, when $k$ grows larger, the difference between the amplitudes of $\ket{0}_p$ and $\ket{1}_p$ becomes smaller. This circumstance increases the difficulty of judging the correct $x_k$. However, in the problem discussed in the present article, it would not matter much: what we have to compute with precision are the real eigenvalues; regarding the complex ones, we have only to detect and discard them by rough estimation.   
\end{itemize}

\section{Summary and discussion} 
 
In this article, we have investigated the feasibility of first-principles molecular dynamics using quantum phase estimation. The algorithm used in this study is composed of the symbolic part and the quantum algorithmic part. In the symbolic part, we construct the set of matrices whose eigenvalues give the root of the governing equation in the given problem. In the quantum algorithmic part, we use the quantum phase estimation to compute the eigenvalues. We have verified that this scheme works well at least in the Hartree-Fock case for simple molecules, wherein we can determine the optimal configuration of the nuclei and the wavefunction simultaneously. 

To conduct more quantitative calculations, the correction to electronic correlation, such as configuration interaction (CI), would be necessary.  This sort of advanced theory is equally represented by polynomials, and our algorithm has the potential to treat it.  In \cite{kikuchi2023molecular}, we discussed the computation of virtual states that would be located above the ground state molecular orbitals. Those virtual orbitals could be encoded in quantum states lying on the qubits.  Following a similar way, we could compute the virtual states, and compose the CI eigenvalue problem, which would be represented by polynomials.  Moreover, for the effective computation of eigenvalues by QPE, we should use recent refined methods of time-evolution \cite{babbush2018encoding, low2019hamiltonian, gilyen2019quantum} and we should improve the techniques to prepare adequate eigenstates for QPE, such as proposed in  \cite{poulin2009preparing, childs2012hamiltonian, berry2015simulating, poulin2017fast, berry2018improved, low2019hamiltonian, ge2019faster, lin2020optimal}. 

To improve the reliability of the algorithm explained in the present article, we should exploit related techniques, such as fault-tolerant quantum computation (FTQC) \cite{kitaev1997quantum, gottesman1998theory}, error correction and quantum code \cite{fowler2012surface,horsman2012surface,bombin2013introduction,higgott2022pymatching,berent2023decoding,wu2023fusion}. In the studies of algebraic quantum codes, the ideas from number theory, algebraic topology, and discrete mathematics are commonly used \cite{chen2015application, degwekar2015extending, breuckmann2021quantum,delfosse2021almost,grassl2021algebraic}. The connection between quantum mechanics, number theory, discrete mathematics, and algebraic geometry, would be an important theme in the future. Such research shall pave the way for a full-fledged FTQC and also provide us with various topics that should be studied by trustworthy FTQC.

We make a short comment on the extension of the algorithm, by which the quantum dynamics of the nuclei could be simulated. In the present article, we have expressed the total energy of the molecule as a multivariate function composed of the atomic coordinate, the wave function, and the orbital energy.  We treated the atomic positions as coordinate points in the context of classical dynamics.  However, the algorithm presented in this article could get over this restriction.  The wavefunction of the molecule could be given by the direct product of the nuclear and electronic parts
\begin{align}
\ket{\Psi}=\ket{\phi_{nuc}}\ket{\psi_{el}}
\end{align}
The total energy in the model treated in this article shall be given by
\begin{align}
E_{tot}
&=\bra{\Psi}\left(-\frac{1}{2\mu}\frac{\partial^2}{\partial^2 R}+H_{el}
(R)\right)\ket{\Psi} \\\nonumber
& =\bra{\phi_{nuc}}\left( -\frac{1}{2\mu} \frac{\partial^2}{\partial^2 R}\right)\ket{\phi_{nuc}}
+\bra{\phi_{nuc}} \left( E_{el}(\psi,R) \right) \ket{\phi_{nuc}}
\end{align}
In the above, $H_{el}(R)$ is the electronic Hamiltonian and $E_{el}(\psi, R)$ is the polynomial representation of the electronic energy. The dynamics related to the atomic coordinate are expressed by an additional kinetic term with a fictitious mass parameter $\mu$. As $E_{el}(\psi,R)$ is the polynomial of $R$, the expectation value $\bra{\phi}_{nuc} \left( E_{el}(\psi,R) \right) \ket{\phi}_{nuc}$ is analytically computed when ${\phi}_{nuc}$ is represented by the Gaussian-type basis set. Then $E_{tot}$ is also the polynomial given by the LCAO coefficients of the nuclei and electron wavefunctions, and one can study it in a similar way that we have demonstrated. We could determine the energetic minimum, or we could draw the {\it orbit} of $R$ when we set $E_{tot}=E_{r}$ with an arbitrary value of the energy.

%We have already known the importance of pre-processing before the quantum computation. The standard method is the Wigner transformation and other methods akin to it. 
 
Note that the present article proposes another way of data preparation before the quantum computation. In chemical problems, one usually adopts the Wigner transformation, or other methods akin to it, to transform the Hamiltonian into a suitable form for the quantum algorithm. In other words, it is always necessary to rewrite the fundamental equations for the sake of the quantum algorithm, and this task requires a symbolic computation.  In the case of the Wigner transformation, the cost for the symbolic computation is small. On the other hand, the method used in the present article requires the symbolic processing of polynomials. Therein the computational costs for generating Gr\"obner basis, in the worst cases, would scale exponentially with the number of variables \cite{bardet2015complexity}. However, we could anticipate that the problem related to the algorithmic complexity of Gr\"obner basis computation might be mitigated by the parallel character of the quantum computations that could be conducted in the vast expansion of data space held on qubits. The authors of the present article hope that the seminal ideas used in our study will flourish in the coming era of fault-tolerant quantum computer.

\section{Data availability}
\label{DataAvailability}
The programs used in this study and the result of the computation are available on the authors' GitHub:

\url{https://github.com/kikuchiichio/20240314}

 \newpage
 \appendix
 
\section{Appendix: profiles of the matrices used in the computation}
 
In this section, the structure of transformation matrices used in the main body of the article ($m_x$, and $m_r$, and $m_e$) are shown by a kind of heat-maps, so that one can see to what extent those matrices are sparse. The entries $A_{ij}$ in the matrix $A$ are indicated by 0 or 1 if it is equal to either of those values; otherwise, they are given by the symbol X. The maximum and the minimum of the entries are also given. The raw numerical data of the matrices are available at \url{https://github.com/kikuchiichio/20240314}.

\begin{verbatim}    
mx:=
|000000X0X0XXX0X0XX0XXX|
|000000X0X0XXX0X0XX0XXX|
|000000X0X0XXX0X0XX0XXX|
|000000X0X0XXX0X0XX0XXX|
|000000X0X0XXX0X0XX0XXX|
|000000X0X0XXX0X0XX0XXX|
|1000000000000000000000|
|000000X0X0XXX0X0XX0XXX|
|0100000000000000000000|
|000000X0X0XXX0X0XX0XXX|
|0010000000000000000000|
|0001000000000000000000|
|0000100000000000000000|
|000000X0X0XXX0X0XX0XXX|
|0000010000000000000000|
|0000001000000000000000|
|0000000100000000000000|
|0000000001000000000000|
|0000000000001000000000|
|0000000000000100000000|
|0000000000000001000000|
|0000000000000000001000|
max(mx)= 258.60441684588295 min(mx)= -72.19094270794328

me:=
|XXXXXX0X0X000X0X00X000|
|XXXXXX0X0X000X0X00X000|
|XXXXXX0X0X000X0X00X000|
|XXXXXX0X0X000X0X00X000|
|XXXXXX0X0X000X0X00X000|
|XXXXXX0X0X000X0X00X000|
|000000X0X0XXX0X0XX0XXX|
|0100000000000000000000|
|000000X0X0XXX0X0XX0XXX|
|0010000000000000000000|
|000000X0X0XXX0X0XX0XXX|
|000000X0X0XXX0X0XX0XXX|
|000000X0X0XXX0X0XX0XXX|
|0000010000000000000000|
|000000X0X0XXX0X0XX0XXX|
|0000000100000000000000|
|0000000010000000000000|
|0000000000100000000000|
|0000000000000100000000|
|0000000000000010000000|
|0000000000000000100000|
|0000000000000000000100|
max(me)= 4115.255552663199 min(me)= -1486.152450941102

mr:=
|XXXXXX0X0X000X0X00X000|
|XXXXXX0X0X000X0X00X000|
|XXXXXX0X0X000X0X00X000|
|XXXXXX0X0X000X0X00X000|
|1000000000000000000000|
|0100000000000000000000|
|000000X0X0XXX0X0XX0XXX|
|0010000000000000000000|
|000000X0X0XXX0X0XX0XXX|
|0001000000000000000000|
|000000X0X0XXX0X0XX0XXX|
|000000X0X0XXX0X0XX0XXX|
|0000001000000000000000|
|0000000100000000000000|
|0000000010000000000000|
|0000000001000000000000|
|0000000000100000000000|
|0000000000010000000000|
|0000000000000001000000|
|0000000000000000100000|
|0000000000000000010000|
|0000000000000000000010|
max(mr)= 3484.3400839743654 min(mr)= -608.2044403624424

\end{verbatim}
\section{
Hartree-Fock Model
}

We assume the equilateral triangle model of H$_3^+$ (with the bond length $R$) and put atomic bases at three centers A, B, and C. 

The trial wave function is given by
\[
\psi(r)=x\cdot\phi(r,A)+y\cdot\phi(r,B)+z\cdot\phi(r,C).
\]
It is associated with the orbital energy $e$ and the coefficients $(x,y,z)$

The atomic orbitals are given by the STO-3G basis set:
\[
\phi(r)=\sum_{i=1}^3 d_i \exp (-b_i r^2).
\]

In that formula, the parameters are given as
\[
b_i= a_i \zeta^2, 
\]
and
\[
d_i= c_i \left( \frac{2b_i}{\pi} \right)^{3/4}
\]
for $i=1, 2, 3$. The numeral data are given in Tables \ref{STO3G} and \ref{STO3GZ}.

\begin{table}[ht]
\begin{center}
\begin{tabular}{lrr}
\toprule
i &    c(i) &    a(i) \\
\midrule
1 &  0.4446 &  0.1098 \\
2 &  0.5353 &  0.4058 \\
3 &  0.1543 &  2.2277 \\
\bottomrule
\end{tabular}
\caption{
The exponents and coefficients of STO-3G of H.
}
\label{STO3G}
\end{center}
\end{table}

\begin{table}[ht]
\begin{center}
\begin{tabular}{lrr}
\toprule
%\mid rule
$\zeta$ &  1.24 \\
\bottomrule
\end{tabular}
\caption{
The scale factor of STO-3G of H.
}
\label{STO3GZ}
\end{center}
\end{table}

We write the atomic orbital centered at the site $X$ as follows:
\[
\phi(r,X)=\phi(r-X)
\]

We sometimes omit the spatial coordinate $r$ for simplicity.

We use the following molecular integrals for every possible combination of atomic bases (indexed by the centers of the orbitals: P, Q, and so on). The summations over those indices are taken over orbital centers $A$, $B$, and $C$ in the molecule.

Overlap integrals:
\[
S_{PQ}=\left(\phi(P)\middle| \phi(Q)\right)
\]

Kinetic integrals:
\[
K_{PQ}=\left(\phi(P)\middle|\ -\frac{1}{2}\nabla^2\ \middle| \phi(Q)\right)
\]

One-center integrals (namely, the nuclear potentials):
\[
V_{PQ,U}=-\left(\phi(P)\middle|\ \frac{1}{\mid r-U\mid }\ \middle| \phi(Q)\right)
\]

Two-center integrals:
\[
[PQ\mid XY]=\int dr_1\ dr_2\ \phi(r_1,P) \phi(r_1,Q)\ \frac{1}{| r_1-r_2| }\ \phi(r_2,X)\phi(r_2,Y)
\]
 
The skeleton part of the Fock matrix: 
\[
H_{PQ}=K_{PQ}+V_{PQ,A}+V_{PQ,B}+V_{PQ,C}
\]

The density matrix:
\[
D=2\begin{pmatrix}
    x^2 & xy &xz\\
    xy  & y^2 &yz\\
    xz  & xy & z^2
\end{pmatrix}
\]

The electron-electron interaction part of the Fock matrix:
\[
G_{PQ}=\sum_{K,L}D_{KL}[PQ\mid KL]-0.5D_{KL}[PL\mid QK]
\]

The Fock matrix:
\[
F_{PQ}=K_{PQ}+G_{PQ}+V_{PQ,A}+V_{PQ,B}+V_{PQ,C}
\]

The total energy (with the normalization condition and the nuclear-nuclear repulsion):

\[
E_{tot}=\frac{1}{2}\sum_{i,j} D_{ij}\cdot(H_{ij}+F_{ij}) -2 e \left(\sum_{i,j} \frac{1}{2} D_{ij}S_{ij}-1\right) +\frac{3}{R}
\]
 \newpage
\input{appendix1}
\bibliographystyle {unsrt}
 \bibliography{biball}
\end{document}

%% file: appendix1.tex
\section{Quantum computation of first-principles molecular dynamics: quest for smaller computational complexity}

In this section, we present a quantum algorithm that enables us to perform first-principles molecular dynamics. In this algorithm, the optimum of wavefunctions and atomic structures is given by a root of a system of polynomial equations derived from the first principles of quantum mechanics. The computational steps consist of symbolic and quantum parts. First, the symbolic computation prepares a set of matrices whose eigenvalues are the roots of the given equation. Second, the quantum computation evaluates the eigenvectors and eigenvalues. In a preceding work, the authors presented another algorithm directed to the same end, and its deficiency is the complexity of the symbolic computations, which would be enormous in the worst cases. In this study, we try to reduce the cost of symbolic computation, utilizing the latest achievements of classical algorithms. Finally, we show how to transform this classical algorithm into a quantum one.    

\subsection{Preliminaries}

The quantum algorithm of first-principles molecular dynamics studied in the main part of this article and a related work  \cite{kikuchi2024extension}, namely, the simultaneous optimization of the electronic and atomic degrees of freedom, is built on an algebraic idea, which is summarized as follows \cite{sottile2002enumerative,moller1995multivariate}. 

\begin{itemize}
\item The governing equations are given by a system of polynomials composed of the following variables: the LCAO coefficients for the wavefunctions, the orbital energies, and the atomic coordinates. We denote the polynomials by $F=\{p_1,p_2, ..., p_M\}$ defined in a ring $\mathcal{R}[x_1,x_2,...,x_n]$. 

\item Compute the Gr\"obner basis $G$ for $F$. 

\item Prepare the monomial basis in the quotient ring $\mathcal{R}[x_1,x_2,...,x_n]/I$, where $I$ is a polynomial ideal generated by $G$.

\item Prepare the numerical matrices that represent the multiplication of each variable with the monomial basis. The eigenvalues and the eigenvectors of those matrices give the roots of $p_1=p_2= ...=p_M=0$.

\item Apply quantum phase estimation to evaluate and encode the eigenvalues, namely, the common roots of the given polynomials in a quantum state.

\end{itemize}

The most demanding part of this algorithm is the computation of Gr\"obner basis since its complexity scales exponentially \cite{bardet2015complexity} with the number of the variables in the worst cases. To the best of the authors' knowledge, no effective quantum algorithm for this issue has yet been developed. 

However, we might reduce the cost of symbolic computations in the quantum algorithm explained above, since there is a useful classical algorithm. It goes without computing Gr\"obner basis and computes the roots of the system of polynomial equations by solving eigenvalue problems \cite{vermeersch2021column}. This algorithm can be transformed into a quantum one since the computation of eigenvalues and eigenvectors can be performed by quantum phase estimation. In the following, we explain this classical algorithm and discuss how to modify it into a quantum algorithm.

\subsection{A classical algorithm to solve the system of polynomial equations}

This algorithm uses the concept of the Macaulay matrix defined for a set of polynomials, which is defined as follows. Let $\{f_1, ..., f_s\}$ be a set of polynomials, each with degree $d_i\ (i = 1,..., s)$. Then the Macaulay matrix of degree $d$, denoted by $M(d)$, is the matrix that contains the coefficients of the polynomials, which is defined as follows:
\begin{align}
\begin{pmatrix}
f_1\\
x_1f_1\\
\vdots\\
x_n f_1\\
x_1^2f_1\\
x_1 x_2 f_1\\
x_2^2 f_1\\
\vdots\\
x_n^{d-d_1}f_1\\
f_2\\
\vdots\\
x_n^{d-d_s}f_s\\
\end{pmatrix}
= M(d) \cdot \hat{X}. 
\end{align}
In the above, each polynomial $f_i$ is multiplied by all monomials with degrees from zero to $d-d_i$, and the resulting polynomials are placed at the column vector on the left-hand side of the relation. $\hat{X}$ is a vector that consists of 1 and the monomials with degrees $\le d$, i.e.,
\begin{align}
    \hat{X}=(1,x_1,...,x_n,x_1^2,x_1x_2,...,x_n^2,...,x_1^d,...,x_n^d)^T.
\end{align}
The dimension of the Macaulay matrix is $r(d) \times q(d)$. The size of the rows and columns are defined by
\begin{align}
r(d)=\sum_{i}^s\begin{pmatrix}d-d_i+n\\ n\end{pmatrix}
\end{align}
and
\begin{align}
q(d)=\sum_{i}^s\begin{pmatrix}d+n\\ n\end{pmatrix}.
\end{align}

%The set of polynomial equations is given by the product of the Macaulay matrix and a vector composed of polynomials:
%\begin{align}
%M(d)\hat{X}=0
%\end{align}
%Hence the null space of the Macaulay matrix gives the common roots of the polynomials. To be precise, the solution vectors consist of the power products of the roots. 

The core of the algorithm is to extract the common roots of $\{f_1, ..., f_s\}$ as eigenvectors and eigenvalues. To this end, we define a series of rectangular matrices representing permutations (i.e., linear transformations) on $\hat{X}$.

\subsubsection{A simple example}
To illustrate the algorithm, we first study a simple two-level model given by
\begin{align}
\begin{pmatrix}0 & -1 \\ -1  & 0\end{pmatrix}
\begin{pmatrix}x \\ y\end{pmatrix}
=
e\begin{pmatrix}x \\ y\end{pmatrix}.
\end{align}

The polynomials involved in this model are as follows:
\begin{align}
ey+x,  ex + y, x^2 + y^2 - 1.
\end{align}

From these polynomials we obtain $M(d)$ and $\hat{X}$ for arbitrary degrees $d$. Then the equations of the two-level model are represented by a set of linear equations:
\begin{align}
M(d)\hat{X}=0.
\end{align}
This relation affirms that we can construct the solutions $(x,y,e)$ out of the column vectors of the null space of $M(d)$. For example, the Macaulay matrix $M(2)$ and $\hat{X}$ at $d=2$ are given by
\begin{align}
M(2)=\begin{pmatrix}
 -1 &  0 &  0 &  0 &  1 &  0 &  0 &  1 &  0 &  0\\
  0 &  1 &  0 &  0 &  0 &  0 &  0 &  0 &  1 &  0\\
  0 &  0 &  1 &  0 &  0 &  0 &  1 &  0 &  0 &  0.
\end{pmatrix}
\end{align}
and
$$
\hat{X}^T=[1, x, y, e, x^2, xy, ex, y^2, ey, e^2]. 
$$

In the classical algorithm proposed in \cite{vermeersch2021column}, the root-finding is performed by a numerical computation of an eigenvalue problem. To this end, we use several matrices that represent the multiplications of variables ($(x, e, R)$) with the monomial vectors $\hat{X}$. Those matrices are defined as follows.

\begin{align}
\begin{pmatrix}
       1\\x\\e\\R
\end{pmatrix}\times{x_g}=S_g 
\begin{pmatrix}
       1\\x\\ y\\ e\\ x^2\\ xy\\ ex\\ y^2 \\ ey\\ e^2
\end{pmatrix}
\ \text{for\ \ }x_g=1,x,e,R.
\end{align}

We obtain the following matrices.
\begin{align}
S_1=
\begin{pmatrix}
       1  &       0  &       0  &       0  &       0  &       0  &       0  &       0  &       0  &       0 \\
       0  &       1  &       0  &       0  &       0  &       0  &       0  &       0  &       0  &       0 \\
       0  &       0  &       1  &       0  &       0  &       0  &       0  &       0  &       0  &       0 \\
       0  &       0  &       0  &       1  &       0  &       0  &       0  &       0  &       0  &       0 
\end{pmatrix}\\
S_x=
\begin{pmatrix}
       0  &       1  &       0  &       0  &       0  &       0  &       0  &       0  &       0  &       0 \\
       0  &       0  &       0  &       0  &       1  &       0  &       0  &       0  &       0  &       0 \\
       0  &       0  &       0  &       0  &       0  &       1  &       0  &       0  &       0  &       0 \\
       0  &       0  &       0  &       0  &       0  &       0  &       1  &       0  &       0  &       0 
\end{pmatrix}\\
S_e=
\begin{pmatrix}
       0  &       0  &       1  &       0  &       0  &       0  &       0  &       0  &       0  &       0 \\
       0  &       0  &       0  &       0  &       0  &       1  &       0  &       0  &       0  &       0 \\
       0  &       0  &       0  &       0  &       0  &       0  &       0  &       1  &       0  &       0 \\
       0  &       0  &       0  &       0  &       0  &       0  &       0  &       0  &       1  &       0 
\end{pmatrix}\\
S_R=
\begin{pmatrix}
       0  &       0  &       0  &       1  &       0  &       0  &       0  &       0  &       0  &       0 \\
       0  &       0  &       0  &       0  &       0  &       0  &       1  &       0  &       0  &       0 \\
       0  &       0  &       0  &       0  &       0  &       0  &       0  &       0  &       1  &       0 \\
       0  &       0  &       0  &       0  &       0  &       0  &       0  &       0  &       0  &       1 
\end{pmatrix}
\end{align}

The matrices defined above satisfy the following relations.

\begin{align}
S_1 \hat{X}\times x_g = S_{x_g} \hat{X} \ \text{for\ }x_g=x,e,R.
\end{align}

Those relations appear to be eigenvalue problems with unknown eigenvector $\hat{X}$, eigenvalues $(x,e,R)$, and concretely defined numerical matrices $S_1$, $S_x$, $S_y$, and $S_Z$. We rewrite it as follows:
\begin{align}
S_1 V D_g = S_{x_g} V \ \text{for\ \ }x_g=x, e, R,
\end{align}
where $V$ are the eigenvectors and $D_g$ is a diagonal matrix that contains eigenvalues. 

In the classical treatment, $V$ is generated by the basis ($Z$) of the null space of the Macaulay matrix through the relation $V=ZT$ with a transformation matrix $T$. Then the eigenvalue problem is transformed into the following form:
\begin{align}
(S_1 Z) T D_g = (S_{x_g} Z)T\ \text{for\ \ }x_g=x, e, R.
\end{align}
In the above, $T$ are eigenvectors and $D_g$ includes the eigenvalues in its diagonal part. However, this equation is generally overdetermined and the matrices on both sides are neither square nor non-singular. We use the Moore-Penrose pseudoinverse of $(S_1)Z$ and modify the problem into a standard form:
\begin{align}
T D_g = (S_1 Z)^\dagger (S_{x_g} Z)T\ \text{for\ }x_g=x, e, R,
\end{align}
whereby we solve the eigenvalue problem of a square matrix $(S_1 Z)^\dagger (S_{x_1} Z)$.

Let us compute the eigenvalue problem defined above by changing $d$ and get the solutions of the two-level model. The result is given in Tables \ref{DataofMacaulayTwoLevelModel} and \ref{SolutionsofMacaulayTwoLevelModel}.

\begin{table}[H]
\begin{center}
\begin{tabular}{ccccc}
       \toprule
        d & Size & Ratio of nonzero elements & Rank & Dimension of computed null space \\
       \midrule
       2 & (3, 10) & 7/30 & 3 & 7 \\
       3 & (12, 20) & 28/240 & 12 & 8 \\
       4 & (30, 35) & 70/1050 & 27 & 8 \\
       8 & (252, 165) & 588/41580 & 157 & 8 \\
       10 & (495, 286) & 1155/141570 & 278 & 8 \\
       \bottomrule
\end{tabular}
\end{center}
\caption{Data of Macaulay matrices for Two-level model with different $d$. This table shows the sizes of $M(d)$, the ratios of non-zero elements in them, their ranks, and the dimensions of the computed null spaces.}
\label{DataofMacaulayTwoLevelModel}
\end{table}

\begin{table}[H]
\begin{center}
\begin{tabular}{c|ccc|ccc}
       \toprule
       d & $x_1$ & $y_1$ & $e_1$ & $x_2$ & $y_2$ & $e_2$ \\
       \midrule
       %2 & No solution &  &  &  &  &  \\
       3 & 0.707107 & -0.707107 & 1.000000 & 0.707107 & 0.707107 & -1.000000 \\
       4 & 0.707107 & -0.707107 & 1.000000 & 0.707107 & 0.707107 & -1.000000 \\
       8 & 0.707107 & -0.707107 & 1.000000 & 0.707107 & 0.707107 & -1.000000 \\
       10 & 0.707107 & -0.707107 & 1.000000 & 0.707107 & 0.707107 & -1.000000 \\
       \bottomrule
\end{tabular}
\end{center}
\caption{Solutions of the two-level model with different $d$. This table shows the two real solutions $(x_1,y_1,e_1)$ and $(x_2,y_2, e_2)$. Note that there are other solutions given by
$(-x_1,-y_1,e_1)$ and $(-x_2,-y_2, e_2)$. At $d=2$, there is no admissible solution, due to the shortage of independent row vectors in the null space of $M(2)$.}
\label{SolutionsofMacaulayTwoLevelModel}
\end{table}

To obtain solutions with accuracy, we must set the degree of the Macaulay matrix large enough. As explained in \cite{vermeersch2021column}, a Macaulay matrix has two types of vectors in its null space: i.e., affine ones and those at infinity.
The affine solutions are defined by
\begin{align}
M(d)\begin{pmatrix} 1 \\ w_1 \\ w_2 \\ \vdots \\ w_N\end{pmatrix}=0.
\end{align}
On the other hand, the solutions at infinity are defined by
\begin{align}
M(d)\begin{pmatrix} 0 \\ w_1 \\ w_2 \\ \vdots \\ w_N\end{pmatrix}=0.
\end{align}
The latter type is not admissible in our problem because the first entry of the solution vector must correspond to the unity, being non-zero.
However, when $d$ is not sufficiently large, it might happen that the obtained solution vector would be a linear combination of affine solutions and those at infinity. It results from the shortage of independent vectors in the computed null space of the Macaulay matrix. An effective way to avoid it is to raise the degree $d$ large enough, whereby the independent solution vectors are generated from the computational basis of the null space without failure.  
%To get reliable solutions, it is sufficient that the number of the independent column vectors in the null space of the Macaulay matrix is as large as the Bezut number $m_b$, as discussed in \cite{}. To have sufficiently large null space, we must set the degree of the Macaulay matrix sufficiently large.  
The key to successful computations is to ascertain a circumstance where the computationally obtained null space contains the column basis vectors as many as the B\'ezout number $m_b=\prod_i d_i$ (the product of the degrees of the polynomials involved in the problem): those independent vectors could produce all of the solutions if there is no numerical error. An idealistic situation is that the dimension of $Z$ stabilizes (or stops growing) with the increase of the degree of the Macaulay matrix. However, due to numerical errors, $Z$ sometimes contains more basis vectors than the B\'ezout number, and it would cause no problem in practice: the more the number of the bases grows, the more accurate the result becomes, as usual with linear computations.

\subsubsection{Example: the optimization of electronic and atomic structure of a simple molecule}
Second, we optimize the electronic and atomic structure of H$_3^+$ from the first principles of quantum mechanics, using the polynomial equations in a similar way as Ref.\cite{kikuchi2023molecular} demonstrated.
The molecule has an equilateral structure made of three hydrogen atoms $A$, $B$, and $C$. The objective function of this model is given by
\begin{Verbatim}[breaklines=True]
f_OBJ=-25940329*R**3*e*x**2 - 61451313*R**3*x**4 + 65640150*R**3*x**2 - 28577961*R**3 + 81961639*R**2*e*x**2 + 1099859207*R**2*x**4 - 811868595*R**2*x**2 + 205761316*R**2 + 342231572*R*e*x**2 - 5233649558*R*x**4 + 3595948148*R*x**2 - 555555556*R - 1960143305*e*x**2 + 200000000*e + 8467967598*x**4 - 6382868964*x**2 + 666666666;
\end{Verbatim}
where $x$ is the coefficient of the wavefunction $\psi(r)=x\phi_{1s}(r_-R_A)+x\phi_{1s}(r_-R_B)+x\phi_{1s}(r_-R_C)$,
$R$ is the interatomic distance of the equilateral triangle structure of H$_3^+$, $e$ the orbital energy, and $R$ the interatomic distance. This objective function is derived from the Taylor expansion of the analytic objective function. The Taylor expansion is applied around ($R_c=1.8$); some solutions with $R$, whose value is too far from $R_c$, would be ineligible.

The system of polynomial equations is defined by
\begin{align}
       p_1&=\frac{\partial f_{OBJ} }{\partial x}=0\\
       p_2&=\frac{\partial f_{OBJ} }{\partial e}=0\\
       p_3&=\frac{\partial f_{OBJ} }{\partial R}=0
\end{align}
The degrees of the polynomials $p_1$, $p_2$, and $p_3$ are 6, 5, and 6, respectively. The system of the polynomial equations has the B\'ezout number 180.

The data of Macaulay matrices and solutions representing the ground state with different $d$ are given in Tables \ref{DataofMacaulayH3} and \ref{SolutionsofH3}. In the computation, the row vectors of the null space are taken from the right singular vectors of $M(d)$ when they have singular values smaller than $0.0001$. At $d<12$, the Macaulay matrices do not have null spaces large enough to compose the solutions of the given equation of the problem. Meanwhile, at $d\ge 12$ the null space becomes large enough to construct solutions. The dimension of the computed null space does not stabilize with the increase of $d$. Instead, the computed null space grows with $d$. This phenomenon is attributed to numerical errors that cause difficulty in distinguishing the correct null space. However, we can compose the solution with substantial exactitude if there are superfluous basis vectors, and we have a criterion to get the proper ground state of the molecule, as mentioned above.

\begin{table}[H]
\begin{center}
\begin{tabular}{ccccc}
       \toprule
       d & Size & Ratio of nonzero elements& Rank  & Dimension of computed null space \\
       \midrule
       6 & (6, 84) & 44/504 & 6 & 78 \\
       8 & (40, 165) & 340/6600 & 40 & 125 \\
       10 & (126, 286) & 1120/36036 & 126 & 160 \\
       12 & (288, 455) & 2616/131040 & 275 & 180 \\
       16 & (936, 969) & 8684/906984 & 749 & 220 \\
       20 & (2176, 1771) & 20400/3853696 & 1511 & 260 \\
       30 & (9126, 5456) & 86580/49791456 & 5096 & 360 \\
       \bottomrule
\end{tabular}
\end{center}
\caption{Data of Macaulay matrices for the equation of H$_3^+$  with different $d$. This table shows the sizes of $M(d)$, the ratios of non-zero elements in them, their ranks, and the dimensions of the computed null spaces.}
\label{DataofMacaulayH3}
\end{table}

\begin{table}[H]
\begin{center}
       \begin{tabular}{cccc}
              \toprule
              d & x & e & R \\
              \midrule
              %6 & No solution &  &  \\
              %8 & No solution &  &  \\
              10 & $\pm$ 0.388666 &-1.204776 & 1.613305\\
              12 & $\pm$ 0.398934 & -1.173845 & 1.743218 \\
              16 & $\pm$ 0.403710 & -1.154682 & 1.807660 \\
              20 & $\pm$ 0.404719 & -1.149509 & 1.823148 \\
              30 & $\pm$ 0.404978 & -1.148190 & 1.827109 \\
              \bottomrule
              \end{tabular}
\end{center}
\caption{Solutions of the equation of H$_3^+$ with different $d$. This table shows one of the real solutions that gives the ground state of the molecule. 
The other solutions that are not presented in the table are irrelevant for various reasons, such as being complex numbers or located outside of the valid range of the model.}
\label{SolutionsofH3}
\end{table}

\subsection{Modification to quantum algorithm}

We use quantum phase estimation to solve the eigenvalue problem. The important point is how to perform the matrix operations on the quantum states in the quantum circuit. What we need for quantum computation is listed below.

\begin{itemize}

\item Operations of arbitrary matrices on statevectors in quantum circuits. It is performed by block-encoding \cite{camps2023explicitquantumcircuitsblock, kuklinski2024sfablelsfablefastapproximate}.

\item The pseudoinverse of matrices. They are prepared by quantum singular value transformation \cite{Gily_n_2019, Martyn_2021, Wang_2021}.

\item The quantum phase estimation. It is performed by various techniques, namely, quantum signal processing,  qubitization, Trotterization, and so on \cite{Low_2019, Yoshioka_2024, baysmidt2025faulttolerantquantumsimulationgeneralized}. Since the problems in this study generally have eigenvalues of complex numbers, the calculations must be carried out in some special ways
\cite{wang2010measurement,daskin2014universal,shao2022computing}.

\end{itemize}

Let us review the classical algorithm used in the previous section and consider how to use quantum routines. It consists of the following steps.

\begin{itemize}
\item Preparation of Macaulay matrices $M(d)$ for the system of polynomial equations defined in $\mathcal{R}[x_1,x_2,...,x_n]$. We use classical computers for this task.

\item Preparation of transformation matrices $S_1$ and $S_{x_i}$ for indeterminate variables $\{x_i\}_i$. We use classical computers for this task.

\item Construction of eigenvalue problems.
%\begin{align}
%       (S_1 Z) T D_g = (S_{x_g} Z)T\ \text{for\ i=1,2,...,n }     
%\end{align}
In classical computation, we represent the eigenvectors by $ZT$ with the null space $Z$ and an unknown matrix $T$. In quantum computation, we could use quantum phase estimation with a statevector $\ket{\psi}$ projected into the null space of $M(d)$, instead of the eigenvectors given by $ZT$. The modified eigenvalue problem is as follows:
\begin{align}
       S_1 Q\ket{\psi} e_g = S_{x_g} Q\ket{\psi}\ \text{for\ i=1,2,...,n. }  
       \label{eqtypeA}
\end{align}
In the above, $Q$ is the projection operator into the null space of $M(d)$. To prepare $Q\ket{\psi}$, it is necessary to compute $M(d)^\dagger M(d) \ket{\psi}$.  We discuss later how to perform this projection in a quantum circuit. 

\item The quantum phase estimation is applied to the square matrix at the right-hand side of the following equation:
\begin{align}
       Q\ket{\psi} e_g = (S_1^\dagger S_{x_g}) Q\ket{\psi}\ \text{for\ i=1,2,...,n. } 
       \label{eqtypeB}
\end{align}
In the above, the pseudoinverse of $S_1$ is just a transposition. %Hence, the operator $A_g$ for which we should evaluate the eigenvalues is given by $A_g = S_1^\dagger (S_g)$ for some $g$, where $g$ corresponds to a variable. 

\item The state after the projection ($Q\ket{\psi}$) is a linear combination of statevectors $\{\hat{X}_i\}_i$,  which are the eigenvectors of the operator $(S_1^\dagger S_{x_g})$:
$Q\ket{\psi}=\sum_i \beta_i {\ket{\hat{X}_i}}$.  We apply the successive quantum phase estimation to $\ket{\psi}\ket{0....}$ to encode the eigenvalues as follows:
$$
Q\ket{\psi}\ket{0....0}\rightarrow\sum_i \beta_i  {\ket{\hat{X}_i}}\ket{E^{(i)}_{x_1},E^{(i)}_{x_2},...,E^{(i)}_{x_n}},
$$
In the above, $\ket{E^{(i)}_{x_1}, E^{(i)}_{x_2},..., E^{(i)}_{x_n}}$ are the ancilla parts where the eigenvalues (namely, the common roots of the initially given polynomials) are recorded.

%We can construct the necessary operators using QSVT and block-encoding.

\end{itemize}

The necessary projection in the quantum algorithm is realized in a quantum circuit in the following way.

\begin{figure}[H]
\begin{center}
    \begin{quantikz}[wire types={q,q,n,q}]      
        \ket{0}    & \gate{H}        & \ctrl{1}      & \gate{H}  &    \meter{}   \\
        \lstick[3]{\ket{\psi}} &                 & \gate[3]{P}              &           &                \\
                    &  \vdots         &  &   \vdots          &                 \\
                   &                 &  &           &                 \\
    \end{quantikz}
\end{center}
\caption{The projector circuit used in the quantum algorithm.}
\label{TheProjectionCircuit}
\end{figure}
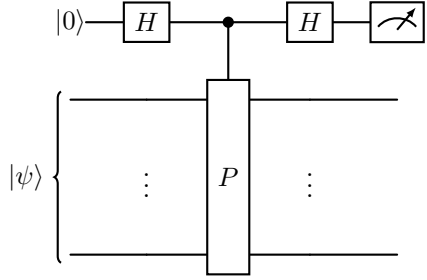

\begin{itemize}
\item First, we must prepare the initial state vector that lies in the null space of the Macaulay matrix $M(d)$. The projector into the null space is given by
\begin{align}
Q=I-M(d)^\dagger M(d)
\end{align}

We perform the following projection of the initial state vector $\ket{\psi}$:
$$\ket{\psi}\rightarrow Q \ket{\psi}$$

\item We need the pseudoinverse of $M(d)$ and we prepare it using quantum singular value decomposition (QSVT).

\item We apply one Hadamard operation at an ancilla qubit to perform the following operation.

$$\ket{\psi}\ket{0} \xrightarrow{H} \frac{1}{\sqrt{2}}(\ket{\psi}\ket{0}+\ket{\psi}\ket{1})$$

\item For simplicity, let $P=I-Q=M(d)^\dagger M(d)$. The operation of $Q$ on the state vector $\ket{\psi}$ is performed by block-encoding. The controlled operation of this operator can be implemented.

\item Then the projection of $\ket{\psi}$ into the null space of $M(d)$ is performed by the following gate operations.

$$
\frac{1}{\sqrt{2}}(\ket{\psi}\ket{0}+\ket{\psi}\ket{1})\xrightarrow{P} \frac{1}{\sqrt{2}}(\ket{\psi}\ket{0}+P\ket{\psi}\ket{1})
$$

$$
\frac{1}{\sqrt{2}}(\ket{\psi}\ket{0}+P\ket{\psi}\ket{1})\xrightarrow{H} \frac{1}{2}((I+P)\ket{\psi}\ket{0}+(I-P)\ket{\psi}\ket{1})
$$

The corresponding quantum circuit is illustrated in Fig. \ref{TheProjectionCircuit}.

\item We measure the control qubit and adopt the component $(I-P)\ket{\psi}\ket{1})$ as the statevectors located in the null space of $M(d)$.

\end{itemize}

Instead of quantum phase estimation, it is possible to use variational approaches to find the solutions of Eq. (\ref{eqtypeA}). Variational approaches adjust the statevectors and eigenvalues by gate operations, with which some controllable parameters are provided. Those approaches have merit, whereby we confine the range of exploration to real numbers by using real statevectors and eigenvalues for trials. 

\subsection{Summary and discussion}
We demonstrated that the root-finding of the set of polynomial equations, which describes a problem of first-principles molecular dynamics, is replaced by a kind of eigenvalue problem in a classical algorithm. To transform the set of polynomial equations into an eigenvalue problem, we used Macaulay matrices and a set of matrices that correspond to the permutations of monomials. From those matrices, we obtain a series of eigenvalue problems for square matrices. We discussed how to modify this classical algorithm into a quantum one, which uses the quantum singular value transformation (QSVT) as an indispensable ingredient for the matrix operations to solve the eigenvalue problem. The use of Macaulay matrices in symbolic computation helps us save computational costs to compose eigenvalue problems. However, the complexity of quantum circuits shall grow due to the complicated operations involved in QSVT. As we saw in the examples in this study, the Macaulay matrices tend to become huge. For example, in the case of the first-principles molecular dynamics of $H_3^+$, we used Macaulay matrices that include thousands of rows and columns, while non-zero elements in them are quite few. It gives rise to the complexity of quantum circuits that require quantum phase estimation, block encoding, and pseudoinverse. On the other hand, it is sufficient to use $22\times 22$ matrices if we prepare eigenvalue problems through Gr\"obner basis \cite{kikuchi2023molecular}. Consequently, the necessary quantum circuit for the latter case is simpler \cite{kikuchi2024feasibility}. Therefore, at present, we should not pass hasty judgment concerning the superiority of these two types of algorithms, which use the Macaulay matrices or Gr\"obner basis, respectively. Instead, we should make a suitable choice between them reflecting the available power of quantum processors when we can use real quantum computers.

\subsection*{Code Availability}
The test programs used in this appendix are available from the authors' GitHub:

\url{https://github.com/kikuchiichio/20250225}